\documentclass{ieee}

\usepackage{setspace}
\usepackage{graphicx}

\begin{document}

\title{Prototype of Fault Adaptive Embedded Software for Large-Scale Real-Time Systems}
\date{}

\author{
Derek Messie, Mina Jung, and Jae C. Oh\\
Department of Electrical Engineering and Computer Science\\
Syracuse University\\
Syracuse, NY 13244\\
jcoh@ecs.syr.edu, dsmessie@syr.edu, mijung@syr.edu\\
\\
Shweta Shetty and Steven Nordstrom\\
Institute for Software Integrated Systems\\
Vanderbilt University\\
Nashville, TN 37235\\
shweta.shetty@vanderbilt.edu, steve.nordstrom@vanderbilt.edu\\
\\
Michael Haney\\
High Energy Physics\\
University of Illinois at Urbana-Champaign\\
Urbana, IL 61801\\
m-haney@uiuc.edu
}

\maketitle
\bibliographystyle{ease05}

\begin{abstract}
This paper describes a comprehensive prototype of large-scale fault adaptive embedded software developed for the proposed Fermilab BTeV high energy physics experiment.  Lightweight self-optimizing agents embedded within Level 1 of the prototype are responsible for proactive and reactive monitoring and mitigation based on specified layers of competence.  The agents are self-protecting, detecting cascading failures using a distributed approach.  Adaptive, reconfigurable, and mobile objects for reliablility are designed to be self-configuring to adapt automatically to dynamically changing environments.  These objects provide a self-healing layer with the ability to discover, diagnose, and react to discontinuities in real-time processing.  A generic modeling environment was developed to facilitate design and implementation of hardware resource specifications, application data flow, and failure mitigation strategies.  Level 1 of the planned BTeV trigger system alone will consist of 2500 DSPs, so the number of components and intractable fault scenarios involved make it impossible to design an `expert system' that applies traditional centralized mitigative strategies based on rules capturing every possible system state.  Instead, a distributed reactive approach is implemented using the tools and methodologies developed by the Real-Time Embedded Systems group.

\end{abstract}

\vspace*{-.3in}
\section{Introduction}
\vspace*{-.05in}
We describe in detail a prototype for the data acquisition and analysis components for the triggering and data acquisition system for the proposed BTeV (http://www.btev.fnal.gov/) system, a particle accelerator-based High Energy Physics (HEP) experiment system at the Fermi National Laboratory. This system consists of a very large number of Digital Signal Processors (DSPs) and General Purpose Processors apart from other hardware components like Field Programmable Gate Arrays (FPGAs) and pixel detectors and sensors. In order to build software for the upcoming BTeV hardware, we formed a Real-Time Embedded Systems (RTES) collaboration, whose responsibility is to develop lower-level real-time embedded intelligent software to ensure system integrity, fault-tolerance, as well as intelligent diagnosis and recovery to process data generated by collisions of physics particles in extremely high data-rate environments (approximately 1.5 Terabytes per second).  Given the complexity of the system, the goal is to develop tools and methodologies that are self-* (self-configuring, self-healing, self-optimizing, self-protecting) as possible.

The BTeV trigger system has three levels, namely Level 1 (L1), Level 2 (L2), and Level 3 (L3). L1 consists of approximately 2,500 DSPs that process data collected from sensors. L2 and L3 are approximately 2,500 Linux machines for processing the data passed through L1 processors. In all three levels, processing the data collected from  sensors is the most important work, which is carried out by High Energy Physics (HEP) applications\footnote{HEP applications are also called {\em physics applications (PAs)}}.  Due to the high-speed data rate and enormous amount of data, the system has to be dynamically fault-adaptive and self-correcting.

Very Lightweight Agents (VLAs) \cite{jo:lightweightagents03} are embedded within L1 as simple software entities which can be implemented in a few dozen lines of assembly language, and take advantage of the exception-signaling and interrupt-handling mechanisms present in most DSP kernels to expose errors in the kernel behavior.

VLAs consist of a {\em proactive part} and a {\em reactive part} to provide fault tolerance in the form of intelligent error detection, diagnosis, and recovery. The proactive part of VLAs can further be divided into a mandatory part and an optional part.  

When the VLA detects (e.g., by monitoring DSP exception signals) an error condition, it may take fault mitigative action directly, or notify appropriate higher level components, which may take appropriate actions such as disabling the execution thread or discarding the current data item.  A similar mechanism will be explored for the monitoring and reporting of deadlines, traffic, processor loads, etc.

The fault tolerance and performance-oriented services offered at L2/L3 will be encapsulated in intelligent active entities (agents) called ARMORs (Adaptive, Reconfigurable, and Mobile Objects for Reliability) \cite{zk:armorover02}. ARMORs are, by design, highly flexible processes, which can be customized to meet the runtime needs of the system. 
 
A prototype for the BTeV L1 trigger system has been built on DSP boards consisting of 16 Texas Instrument DSPs. The prototype includes L1 VLAs, ARMORs, and the Experimental Physics and Industrial Control System (EPICS).  It exhibits several fault adaptiveness and tolerance behaviors. The prototype provided us a great opportunity to realize the ideas and concepts to a real-working hardware platform. This paper describes the design and development of the prototype in detail.

The rest of this paper is organized as follows.  Section 2 provides some background on the BTeV experiment and the RTES collaboration.  The RTES system development environment is then presented in Section 3, including an overview of VLAs and ARMOR.  The various system modeling tools developed within the collaboration, along with an explanation of how each is used for design and implementation is also detailed.  

Section 4 describes the prototype that was presented at SuperComputing 2003 (SC2003).  Design motivation is discussed, followed by software and hardware specifications.  The embedded VLA design and implementation for the prototype is detailed, along with an explanation of the Experimental Physics and Industrial Control System (EPICS) used to inject system faults and monitor VLA mitigation and overall system behavior.  Lessons learned are also provided.

Finally, future efforts planned for the next phase of prototype development are described, followed by a conclusion.

\vspace*{-.05in}
\section{Background and Motivation}
\vspace*{-.05in}
\subsection{RTES/BTeV}
\vspace*{-.05in}
\begin{figure}[b]
\centering
\includegraphics[width=78mm]{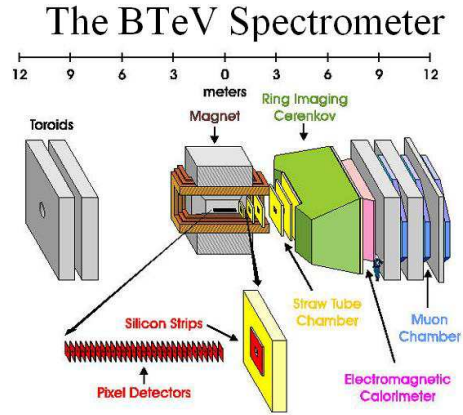}
\caption{BTeV pixel detector layout.}
\label{fig:detectorlayout}
\end{figure}

BTeV is a proposed particle accelerator-based High Energy Physics (HEP) experiment currently under development at Fermi National Accelerator Laboratory.  The goal is to study charge-parity violation, mixing, and rare decays of particles known as beauty and charm hadrons, in order to learn more about matter-antimatter asymmetries that exist in the universe today \cite{sk:pixeldetect02}.  When approved, the BTeV experiment will be sponsored by the Department of Energy.

The BTeV experiment will operate in conjunction with a particle accelerator where the collision of protons with anti-protons can be recorded and examined for detached secondary vertices from charm and beauty hadron decays \cite{eg:detachedtrigger00}.  The layout for the BTeV detector is shown in Figure \ref{fig:detectorlayout}.  

The experiment uses approximately 30 planar silicon pixel detectors to record interactions between colliding protons and antiprotons in the presence of a large magnetic field.  The pixel detectors, along with readout sensors are embedded in the accelerator, which are connected to specialized field-programmable gate arrays (FPGAs).  The FPGAs are connected to approximately 2,500 digital signal processors (DSPs).   

The measurements of the interactions resulting from the collision of protons and antiprotons are carried via custom circuitry hardware to localized processors that reconstruct the 3-dimensional crossing data from the silicon detector in order to examine the trajectories for detached secondary vertices \cite{sn:vandy03}.  These detached vertices are indicators of the likely presence of beauty or charm decays.

BTeV will operate at a luminosity of 2x10$^{32}$cm$^{-2}$s$^{-1}$ corresponding to about 6 interactions per 2.53 MHz beam crossing rate \cite{sk:pixeldetect02}.  Average event sizes will be around 200 Kilobytes after zero-suppression of data is performed on-the-fly by front-end detector electronics.  Every beam crossing will be processed, which translates into the extremely high data rate of approximately 1.5 Terabytes of data every second, from a total of 20x10$^{6}$ data channels.

A three tier hierarchical trigger architecture will be used to handle this high rate.  Data from the pixel detector and muon detector will be sent to the Level 1 trigger processors, where an accept or reject decision will be made for every crossing.  The Level 1 vertex trigger processor will perform pattern recognition, track, and vertex reconstruction on the pixel data for every interaction \cite{sk:pixeldetect02}.  It has been estimated that 99\% of all minimum-bias events will be rejected by the Level 1 vertex trigger, while 60-70\% of the events containing beauty or charm decay will still be accepted for further evaluation.

Levels 2 and 3 will be implemented on a cluster, and data that makes it past the Level 1 filter is assigned to one of these Level 2/3 processors for further analysis.  Data that survives Level 2 will be passed to Level 3 algorithms to determine whether or not it should be recorded on archival media \cite{jb:ftissues02}.

It is estimated that Level 2 will decrease the data rate by a factor of 10, and Level 3 will further reduce the incoming rate by a factor of 2.  Once data is filtered through all three levels, and additional data compression is performed, it is expected that the resulting data rate will be approximately 200 Megabytes per second.

The events that are actually accepted within this system occur very infrequently, and the cost of operating this environment is high.  The extremely large streams of data resulting from the BTeV environment must be processed real-time with highly resilient adaptive fault tolerant systems.  For these reasons, a Real-Time Embedded Systems Collaboration (RTES) was formed with the purpose of designing real-time embedded intelligent software to ensure data integrity and fault-tolerance within this data acquisition system.  The collaboration includes team members from Fermi Lab, Syracuse University, Vanderbilt University, University of Illinois at Urbana-Champaign, and the University of Pittsburgh.

\vspace*{-.05in}
\section{RTES System Development Environment}
\vspace*{-.1in}
\begin{figure*}[t]
\centering
\includegraphics[width=155mm]{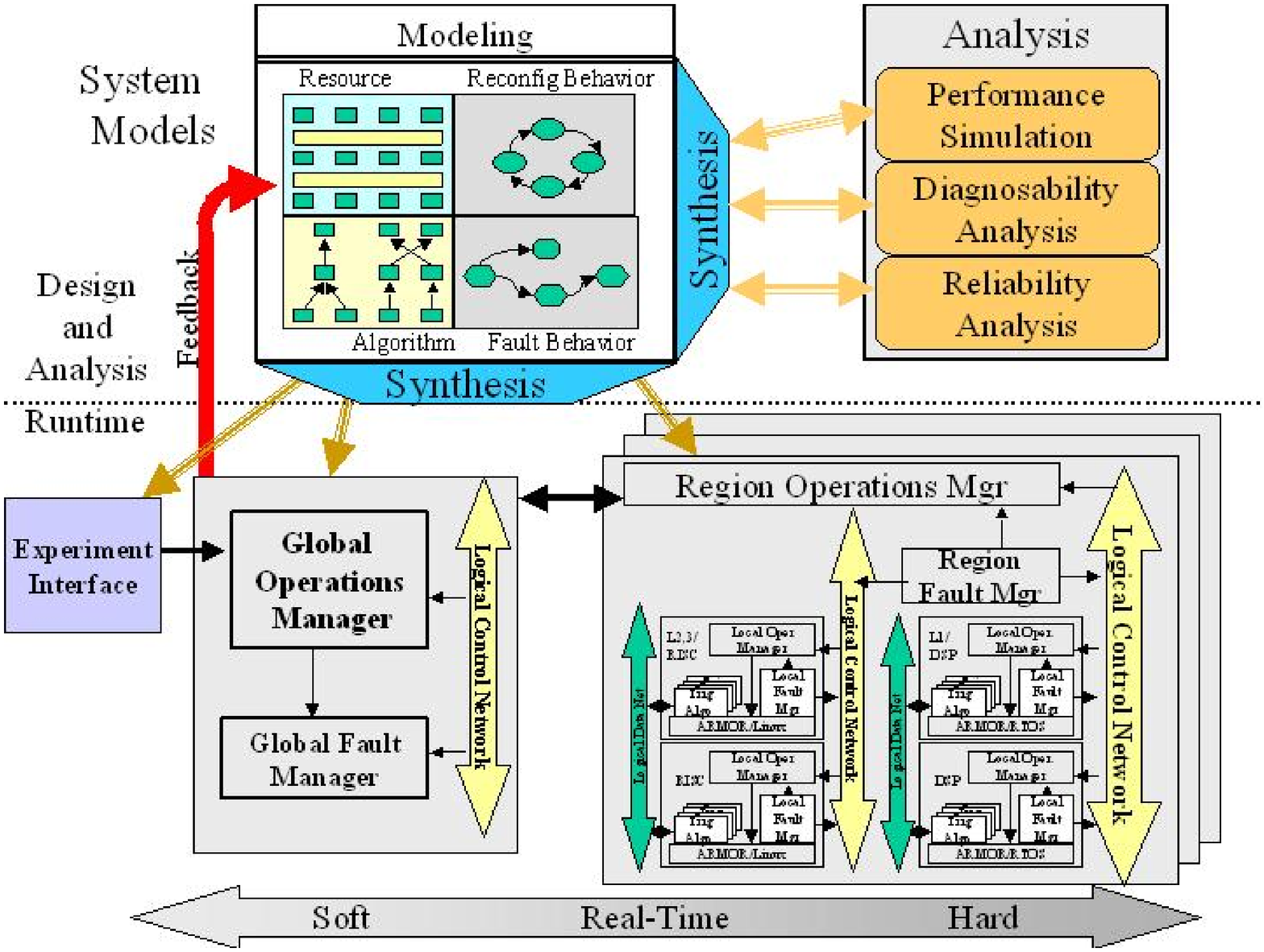}
\caption{\textit{BTeV System Design and Runtime Framework - System Models use domain-specific, multi-view representation formalisms to define system behavior, function, performance, fault interactions, and target hardware.  Analysis tools evaluate predicted performance to guide designers prior to system implementation.  Synthesis tools generate system configurations directly from the models.  A fault-detecting failure-mitigating runtime environment executes these configurations in a real-time, high performance, distributed, heterogeneous target platform, with built-in, model-configured fault mitigation.  Local, regional, and global perspectives are indicated.  On-line cooperation between runtime and modeling/synthesis environment permits global system reconfiguration in extreme-failure conditions.}}
\label{fig:sysdev}
\end{figure*}

An overview of the BTeV system design and runtime framework is shown in Figure \ref{fig:sysdev}.  There are four primary components, including very lightweight embedded fault mitigation agents (VLAs), adaptive, reconfigurable, mobile objects for reliability (ARMOR), a generic modeling environment (GME), and a system operator interface (EPICS).  

EPICS (http://www.aps.anl.gov/epics) provides an interface for injecting faults into the system, which allows for evaluation of the effect of individual fault scenarios on the BTeV environment.  It provides a way for operators to monitor and control overall system behavior.  Details and screenshots of the EPICS interface are presented in later sections of this paper that describe the system prototype.

\vspace*{-.05in}
\subsection{Very Lightweight Agents}
\vspace*{-.05in}
\subsubsection{Overview}
Multiple levels of very lightweight agents (VLAs) are one of the primary components responsible for fault mitigation across Level 1 of the real-time embedded RTES/BTeV data acquisition system \cite{jo:lightweightagents03}.  As described earlier, Level 1 alone is made up of 2,500 DSPs, with each DSP consisting of three components, namely a physics application (PA), a very lightweight agent (VLA), and the DSP kernel itself.  The PA is responsible for running Level 1 data filtering algorithms, while the VLA provides each PA (and the DSP kernel), with a lightweight, adaptive layer of fault mitigation.  Also, several DSPs are grouped onto a single \textit{farmlet} are each assigned a farmlet VLA (FVLA) that provides a layer of fault mitigation across all DSPs within a given farmlet.  Likewise, a regional VLA (RVLA) provides a higher layer of fault mitigation across a group of farmlets.  One of the latest phases of work at Syracuse University has involved implementing individual proactive and reactive rules across multiple layers of VLAs for specific Level 1 system failure scenarios.  

One of the major challenges is to find out how the behavior of the various layers of VLAs will scale when implemented across the 2,500 DSPs projected for Level 1 of the BTeV environment.  In particular, how will rules within each VLA interact as they are activated in parallel at multiple layers of the system, and how will this affect other components and the overall behavior of a large-scale, real-time embedded system such as BTeV.  Given the number of components and countless fault scenarios involved, it would be impossible to design an `expert system' that applies mitigative actions triggered from a central location acting on rules capturing every possible system state.  Rather, the Syracuse team has implemented specific \textit{layers} of fault mitigative behavior similar to Rodney Brooks' multi-layer, decentralized subsumption approach for mobile robot design.

\subsubsection{VLA Subsumption Model}
The phrase \textit{subsumption architecture} was first used by Brooks to describe a bottom-up approach for mobile robot design that relies on multiple layers of distributed sensors for determining actions \cite{rb:brooks86}.  Until that time, designs relied heavily on a centralized location where most, if not all, of the decision making process took place.  In fact, only initial sensor perception and motor control were left to distributed components.  As a result, the success and adaptability of these systems was almost entirely dependent on the accuracy of the model and actions represented within the central location.

In contrast, Brooks proposed that there should be essentially no central control.  Rather, there should be independent layers each made up of a large number of sensors, with each layer responsible for distinct behavior.  Communication and representation is developed in the form of action and inaction at each of the individual layers, with certain layers subsuming other layers when necessary.  In this way, layer after layer is added to achieve what Brooks refers to as \textit{increasing levels of competence}.   This breaks the problem down into \textit{desired external manifestations}, as opposed to slicing the problem on the basis of internal workings of the solution as was typically done in the past \cite{rb:brooks91}.

Multiple layers of individual proactive and reactive VLAs have been embedded within the RTES/BTeV environment.  Lower level \textit{worker} VLAs are responsible for mitigative actions performed at local worker DSPs, while VLAs at higher levels (FVLAs, RVLAs) perform fault mitigation related to components at the farmlet and region level.  In addition, farmlet VLAs monitor and communicate with lower level groups of workers, and may subsume the actions of individual worker VLAs if a pattern of behavior is observed across other workers within the same farmlet.  Similarly, regional VLAs are responsible for fault mitigation at the regional level, and monitor and communicate with lower level farmlet VLAs.  At each layer, individual VLAs are responsible for monitoring and communicating with a specific group of lower level VLAs, and may subsume certain actions if a particular pattern of behavior across the group exists.

\vspace*{-.05in}
\subsection{Adaptive, Reconfigurable, Mobile Objects for Reliability (ARMOR)}
\vspace*{-.05in}
\begin{figure}[b]
\centering
\includegraphics[width=78mm]{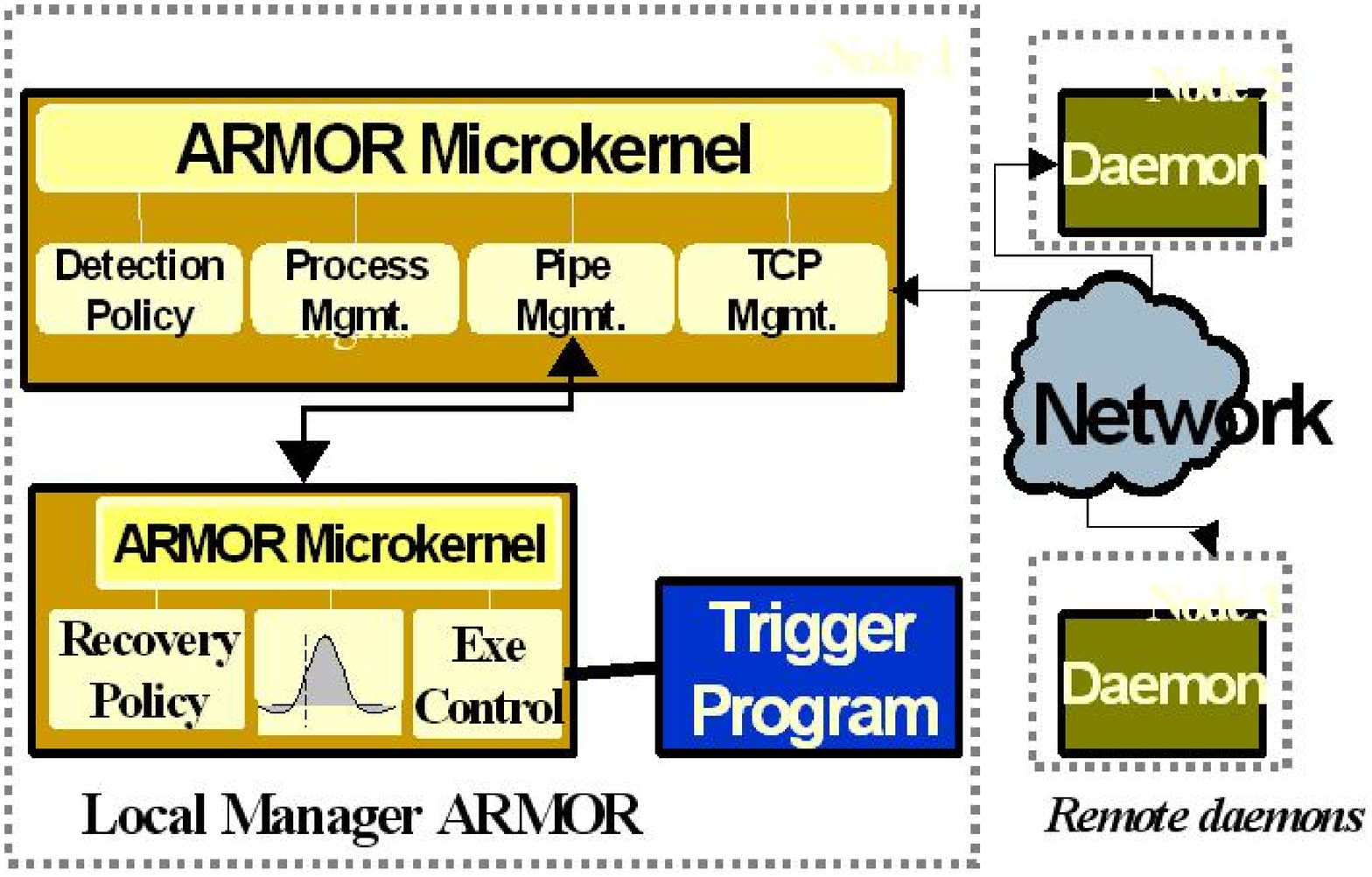}
\vspace*{-.5in}
\caption{Sample ARMOR consisting of multiple Elements}
\label{fig:samplearmor}
\end{figure}

While embedded VLAs provide a lightweight, adaptive layer of fault mitigation within Level 1, the University of Illinois is developing software components that run as multithreaded processes responsible for monitoring and fault mitigation at the process and application layer.

Adaptive, Reconfigurable, and Mobile Objects for Reliability (ARMOR) \cite{zk:armorover02} are multithreaded processes composed of replaceable building blocks called \textit{Elements} that use a messaging system to communicate.  The components within the flexible architecture are designed such that distinct modules responsible for a unique set of tasks can be plugged into the system.  There are separate Elements that are responsible for recovery action, error analysis, and problem detection, which can each be developed and configured independently.  ARMORs are configured in a hierarchy across multiple nodes of the entire system.  A sample ARMOR is shown in Figure \ref{fig:samplearmor}.  In this example, a primary ARMOR daemon is watching over the node and reporting to higher-level ARMORs out on the network \cite{jk:hardwarefailure03}.  Elements within node-level ARMORs communicate to ensure that all nodes are operating properly.

\textit{Execution ARMOR} is responsible for monitoring and ensuring the integrity of a single application, without requiring any modifications to the program itself.  It watches the program to ensure that it is continues to run, and has the ability to restart the application when necessary.  As it is monitoring, it may generate messages for other Elements to analyze and act on based on what it finds.  The Execution ARMOR is also capable of triggering specific recovery actions based on the pattern of return codes that it receives from the application.  Another distinct ARMOR known as \textit{Recovery ARMOR} consists of Elements that have the ability to automatically migrate processes from one machine to another when the work load across machines is not balanced.

Within the trigger, ARMORs provide error detection and recovery services to the trigger system, along with any other processes running on Levels 2 and 3.  Hardware failures may also be detected.  ARMOR components are designed to run under an operating system such as Linux and Windows, and not within low level embedded systems that require real-time memory and processing time constraints.  

There is also an ARMOR API that allows trigger applications to proactively send specific error information directly to an Element.  Data processing and quality rates can also be sent directly to the ARMOR where they may be distributed to corresponding Elements for analysis \cite{jk:hardwarefailure03}.

\vspace*{-.05in}
\subsection{System Modeling Tools}
\vspace*{-.05in}
The Generic Modeling Environment (GME) tool \cite{al:gme00}\cite{tb:vandymodel02} developed by the Institute for Software Integrated Systems (ISIS) at Vanderbilt University provides a graphical language that is used to specify and design the RTES/BTeV environment.  Various aspects of the system can be modeled, including application data flow, hardware resource specification, and failure mitigation strategies.  

The GME tool was used to model several aspects of the prototype described in detail in Section 4.  The \textit{Data Flow Specification Language} was used to specify data flow within the prototype, while the \textit{Resource Specification Language} defined the physical hardware layout.  Portions of the \textit{Fault Mitigation Language} were used as well.

\subsubsection{Data Flow Specification Language}

The application data flow model allows a system developer to define the key software components and the flow of data between them \cite{ss:vandymodel04}.  Standard hierarchical dataflow notation is used, where nodes capture the software components, and connectors show the flow of data between nodes.  These models can represent synchronous or asynchronous behavior, and a variety of scheduling policies.  For the BTeV trigger, these are primarily asynchronous operations, with data-triggered scheduling.

The primitive software components in the dataflow model are associated with a script that provides the implementation of the software component \cite{ss:vandymodel04}.  Fault-manager processes are associated with specific fault-mitigation strategies.

\subsubsection{Resource Specification Language}

A resource specification language defines the physical structure of the target architecture.  Block diagrams capture the processing nodes (CPUs, DSPs, FPGAs), and connections capture the networks and busses over which data can flow.  One of the assumptions made here is that the hardware component is modeled exactly the same way as it is laid out physically. 

\subsubsection{Fault Mitigation Language}

The modeling environment also provides a language for specifying fault mitigation strategies to address hardware resource and data flow failure scenarios.  Statechart-like notation \cite{dh:statecharts87} is used for defining various failure states.  Conditions to enter or leave those states, along with actions to be performed when state transitions occur are also defined \cite{ss:vandymodel04}.  

System states are represented with distinct nodes in the state diagram, each corresponding to a particular phase of system operation.  Lines are used to represent transitions between states, capturing the logical progression of system modes.  Transitions occur when specific events or sets of events are triggered (hardware faults, OS faults, user-defined errors, fault-mitigation commands from higher level VLAs, etc.).  Actions are defined for each trigger (moving tasks, rerouting communications, resetting and validating hardware, changing application algorithms, etc.).

UML is used to capture the various associations and interactions between components in the meta-model for the state machine \cite{tb:umlmodel00}.  Behavior state machines perform actions based on triggering conditions, where a trigger is defined as a connection which contains attributes that define triggering conditions and actions performed.  Statecharts can be used to describe the behavior of individual fault managers.  Ports can be designated as \textit{Input} or \textit{Output}.

\textit{Actions} are written in C, and typically involve forwarding messages upstream or downstream to notify the appropriate layer of VLAs or other necessary system components.  There are three primary types of messages, all of which are passed asynchronously.  \textit{Fault/Error} messages report errors in hardware or the application, while \textit{control} messages are decision requests or commands that force parameters to change in the running system.  The model also allows for defining periodic \textit{statistical} messages.

The overall objective for modeling fault mitigation strategies is to realize minimal functionality loss for any set of possible component failures, recover from failures as quickly and completely as possible, and to minimize the cost associated with excessive hardware redundancy.  

\subsubsection{System Generation}

The overall system is generated automatically once the model has been sufficiently defined by the user.  Several low-level artifacts need to be generated from the models in order to derive an implementation.  System dataflow synthesis involves mapping a specific dataflow model into a set of software processes and inter-process communication paths \cite{ss:vandymodel04}.  This mapping also needs to derive the execution order or schedule of the processes executing on the processors.  The communication paths between software processes must be setup such that the software process itself is unaware of the location of other software processes that it is communicating with.  However, the mapping process alone cannot enforce location-transparent communication.  It relies on some capabilities in the runtime execution infrastructure in order to facilitate this \cite{js:isis00}.

\vspace*{-.05in}
\section{Prototype}
\vspace*{-.05in}

The RTES group has developed various methodologies and tools for designing and implementing very large-scale real-time embedded computer systems.  However, prior to this prototype, there was no single integrated system that had been developed using all of these tools together to meet a common objective.  The overall goal of the prototype was to demonstrate an implementation of these various tools and methodologies within a single system capable of efficient fault mitigation for a set of error conditions.  The prototype was demonstrated at SuperComputing 2003 (SC2003).

\vspace*{-.05in}
\subsection{Component Details}
\begin{figure}[t]
\centering
\includegraphics[width=78mm]{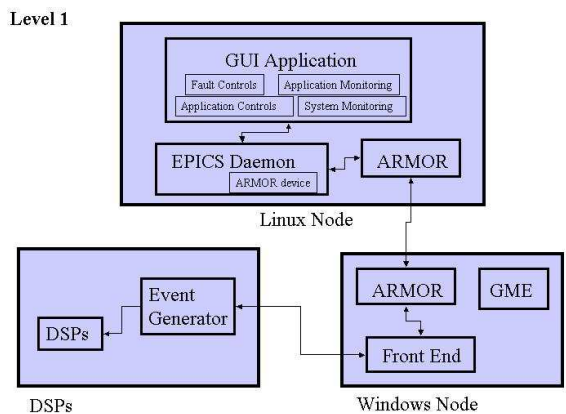}
\caption{Prototype Architecture Overview}
\label{fig:sc03arch}
\end{figure}

\vspace*{-.05in}
An overview of the complete prototype architecture presented at SC2003 is found in Figure \ref{fig:sc03arch}.  Level 1 of the BTeV event filter is the primary setting for the demonstration.  The prototype hardware consists of a 7-slot VME crate with 4 fully populated motherboards and 16 DSPs.  The DSPs were Texas Instruments C6711 with 64MB of RAM each, running at 166 MHz.

\subsubsection{EPICS}
\begin{figure*}[t]
\centering
\hspace*{.3in}
\includegraphics[width=200mm]{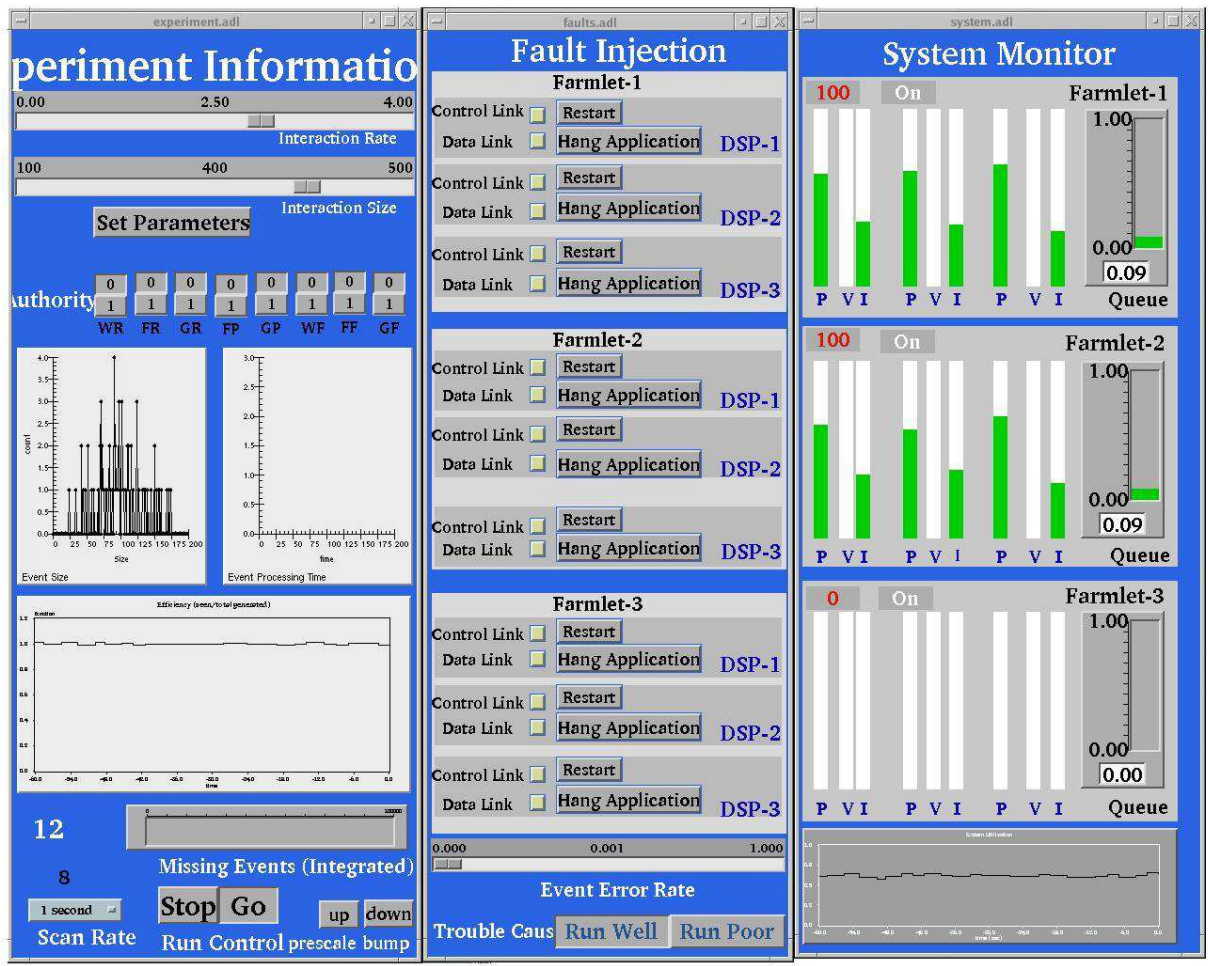}
\vspace*{-.9in}
\caption{EPICS Prototype Screenshot}
\label{fig:protoepics}
\end{figure*}

As mentioned earlier, the Experimental Physics and Industrial Control System (EPICS) was used to provide an interface for controlling the operation of the prototype system, as well as for injecting faults into the system.  A screenshot of the EPICS-controlled prototype presented at SC2003 is shown in Figure \ref{fig:protoepics}.

The left panel titled \textit{Experiment Information} provides a number of controls and display relevant to the BTeV experiment. The \textit{Interaction Rate} and \textit{Interaction Size} sliders affect the generation of physics data; their influence is shown in the \textit{Rate} and \textit{Size} histograms in the middle of the panel. The \textit{Set Parameters} button causes the rate and size sliders to take effect. (The \textit{Authority} buttons will be explained later). The \textit{Efficiency} graph indicates the ratio of processed (not lost) to generated data, and the \textit{Missing Events} displays (number and graph) indicate in absolute terms the events that have been lost. The primary BTeV operator controls (\textit{Stop, Go}) are at the bottom of the panel.

The right panel titled \textit{System Monitor} shows the operational state of the DSPs in each of 3 farmlets. For each DSP, utilization is subdivided into P (physics application), V (VLA), and I (idle) time bar graphs. The \textit{System Monitor} panel also shows the Buffer Manager queue occupancies and overall system utilization. Under normal operation, only 2 farmlets are active; the third is a \textit{hot spare} available to take on work if one of the other farmlets fails.

The center panel titled \textit{Fault Injection} allows the prototype user (not to be confused with the "BTeV operator") to hang or restart the physics application on any of the individual DSPs within a given farmlet, as well as severing the data and control links. An \textit{Error Rate} slider is provided for automatically generating corrupt data, and a \textit{Run Well, Run Poor} button-pair selects whether the physics application reacts gracefully (ignore) or ungracefully (hangs) in response to corrupt events.

There are two `exceptions' with respect to the organization of the controls, reflecting the abstract distinction between the prototype user (someone who would use this prototype) and the BTeV operator (some who would use BTeV). It is unlikely that the BTeV operator would (or would be able to) hang the physics application, but it may well be the case that the operator would have a control to restart the application.  The \textit{Hang} and \textit{Restart} buttons are both shown on the middle panel as user controls, even though \textit{Restart} may be an operator control.

The other exception is the \textit{Authority} control near the top of the left panel. This collection of buttons determines which mitigation strategy(ies) are enabled. While several controls are provided, the most important are:
\begin{itemize}
\item\textbf{Worker Reset (\textit{WR})} - authorizes the VLA on a worker to restart the physics application if it fails to meet a timeout deadline.
\item\textbf{Farmlet Prescale (\textit{FP})} - authorizes a farmlet to determine a farmlet-wide rate for dropping events without analysis, in an effort to prevent queue overflow.
\item\textbf{Global Prescale (\textit{GP})} - similar to farmlet prescale, but the drop rate is uniform across all farmlets.
\item\textbf{Global Failover (\textit{GF})} - authorizes an upper-layer ARMOR to declare a farmlet to be unfit, and to redirect future work to a \textit{hot spare} farmlet.
\end{itemize}

\subsubsection{VLA Prototype}
Multiple layers of proactive and reactive VLAs were implemented within the SC03 prototype.  
Since the physics application (PA) at the worker level is responsible for the critical overall objective of Level 1 data filtering, it is extremely important that DSP usage by the VLA at the worker level is minimal, and only occurs either when the PA is not utilizing the DSP, or when emergency fault mitigative action is required.  For this reason, the prototype worker VLA is implemented as an Interupt Service Routine (ISR) that is triggered only when expected PA processing time thresholds are exceeded.  The TI T6711 DSP processor used within the prototype has 15 hardware interrupts (HWIs).  HWI 15 is assigned Timer 1, and HWI 14 is assigned Timer 0.  The VLA prototype uses HWI 15 (Timer 1). 

One of the fault scenarios modeled within the prototype occurs when the DSP is found to be over the estimated time budget on crossing processing (e1).  In this scenario, HWI 15 (Timer 1) is used by the VLA to monitor PA crossing processing times, and trigger the VLA ISR if the time threshold is exceeded.  At the start of processing each crossing, the PA provides the VLA with a time estimate as to the maximum time that it should take to process the current crossing.  The Timer 1 Period Register (T1PR) is assigned this estimated value, and timer counting is enabled.  If the PA completes crossing processing as expected prior to the timer expiring, then the timer is stopped and reset when the PA begins processing the next crossing.  If on the other hand, the timer expires before the PA has completed processing, then the VLA ISR is called.  The first time that the VLA ISR is triggered, the VLA notifies the PA of the time threshold violation, and resets the timer for a set grace period.  The PA then attempts to cleanup any remaining processing that it has to complete.  If successful, the PA stops the timer, and continues on to the next crossing.  If the cleanup is unsuccessful, the VLA ISR is again called, and this time, it either attempts to reset the PA itself (if it has authority), or sends communication up to the next level of VLA (in this case the Farmlet VLA) for remedial action.

In addition to taking direct fault mitigative actions on various system components, multiple layers of VLAs are also responsible for communicating specific error messages to higher layers within the system.  As detailed in Section 3.1.2 describing the VLA subsumption model, trends in the type and frequency of messages sent to higher level VLAs can lead to subsuming the actions of lower level VLAs.

\vspace*{-.05in}
\subsection{Lessons Learned}
\vspace*{-.05in}
Following the SC2003 conference, a formal review \cite{mf:sc03review} of the RTES/BTeV prototype was conducted by a team consisting of members both internal and external to the project.  Everyone was in agreement as to the substantial value of successfully producing a single integrated system using many of the component designs and tools developed across the collaboration.  There were of course also some valuable lessons learned.  A few of the primary areas of concern cited in the review follow.

Firstly, GME is an integral piece of the software development cycle.  Many different groups within the RTES collaboration will be developing, testing, and releasing various BTeV modules in parallel.  Therefore the review stressed the need to break the current GME model down into sub-models, so that work on distinct subsections of the model can occur simultaneously.  Submodels accessed through a standard change control tool will ease the future coordination and tracking of overall model changes.

Another related issue that was raised in the review is that of overall software release versioning.  Currently, various components and tools for the system are being developed in parallel by different teams within the collaboration.  Since the primary goal is to be able to provide a total integrated \textit{package} of these components and tools, a versioning system must be developed that facilitates a single production version of the BTeV software.  This will make it easier for different developers to work with different versions of distinct components or tools without colliding with each other or with the production release.  BTeV will need to have several production versions in use at one time, and also an arbitrarily large number of development versions.  Establishing a formal versioning system now that spans development efforts will ensure a deliverable of a single integrated BTeV \textit{package}, where consistent versions can be used across multiple development and production environments.

Next, \textit{Elements} of the ARMOR are written in \textbf{Chameleon}, a framework built to be a research vehicle for exploring the world of conceptual programming.  However, since BTeV authors will be expected to invent and implement new Elements quickly, the use of a more standard development language such as \textbf{Python} would help reduce the learning curve and effort required for adding and testing new elements.

Finally, it is critical that physicists that use this system are provided as much detail as possible on tracking changes that occur within the trigger system.  For example, if a prescale value is changed, a log must be kept that allows the user to identify the precise time of the change in order to compare it against the modified behavior experienced within the system.  There must be a standard and easy way to log and review any and all system control changes across time, not just a real-time current view of the values from the graphical user interface.

A formal document \cite{lw:sc03response} in response to the issues raised within the review was also completed.

\vspace*{-.07in}
\section{Next Steps}
\vspace*{-.09in}
As described in the previous section, the design and implementation of the SC2003 prototype was an important step for RTES in showing the integration of many of the component designs and tools that have been developed across the collaboration.  Each of the teams within the collaboration have been able to take away some valuable \textit{lessons learned} that will be incorporated into the development process moving forward.  In addition to addressing these lessons, there are many other challenging goals that RTES has set.

Firstly, as detailed earlier, the prototype included 16 DSPs at L1.  Since the hardware projected for L1 consists of 2,500 such DSPs, RTES needs to demonstrate how the components and tools developed will scale when implemented on a much higher volume of DSPs.  Issues of scalability are one of the primary areas that RTES will be focusing on for the future, and plans are already being made for the next phase of a prototype that will include far more processors and supporting hardware.

Next, VLA research is exploring ways that the lightweight, adaptive nature of the VLA may be further used to coordinate communication and mitigative actions across the large-scale environment.  Adaptive agent architectures that facilitate large-scale coordination are being evaluated for idioms that may address specific challenges within the RTES environment.

The next phase of modeling tools are also being developed that will further support component design and implementation.

\vspace*{-.1in}
\section{Conclusion}
\vspace*{-.08in}
This paper has described a large-scale fault adaptive embedded software prototype for the proposed Fermilab BTeV high energy physics experiment.  Self-optimizing, self-protecting, proactive and reactive Very Lightweight Agents (VLAs) are embedded within Level 1 to provide an adaptive layer of fault mitigation across the RTES/BTeV environment.  Adaptive, Reconfigurable, and Mobile Objects for Reliability (ARMOR) are designed to be self-configuring to adapt automatically to dynamically changing environmental demands.  The prototype demonstrates the self-healing qualities of these objects designed with the ability to discover, diagnose, and react to discontinuities in real-time processing. 

The prototype was developed by the RTES collaboration, whose responsibility is to develop low-level real-time embedded intelligent software to ensure system integrity and fault-tolerance across extremely high data-rate environments.  The objective of the prototype was to produce a single integrated system using many of the component designs and tools developed thus far across the collaboration. 

The Generic Modeling Environment (GME) developed by the Institute for Software Integrated Systems (ISIS) at Vanderbilt University was used to design and implement application data flow, hardware resource specifications, and failure mitigation strategies.  The Experimental Physics and Industrial Control System (EPICS) that was used within the prototype to inject system faults and monitor VLA mitigation and overall system behavior was also presented.

Finally, lessons learned from designing, implementing, and presenting the prototype, along with planned future efforts for the RTES collaboration were also provided.

\vspace*{-.07in}
\textbf{\\
ACKNOWLEDGEMENTS}\\
\vspace*{-.15in}
\\
The research conducted was sponsored by the National Science Foundation in conjunction with Fermi National Laboratories, under the BTeV Project, and in association with RTES, the Real-time, Embedded Systems Group (NSF grant \# ACI-0121658).

\vspace*{-.1in}
\bibliography{ease05}

\end{document}